\documentclass[epj]{svjour}
\usepackage{latexsym}
\usepackage{graphicx}
\usepackage{amssymb}
\usepackage{multirow}
\usepackage{amsmath}
\usepackage{epstopdf}
\begin{document}
\title{Persistent spin splitting of a two-dimensional electron gas in tilted magnetic fields}
\author{Rayda Gammag \and Cristine Villagonzalo}
%
%
\institute{National Institute of Physics, University of the Philippines, Diliman, Quezon City, Philippines 1101}
\date{Received: date / Revised version: date}
%
\abstract{
By varying the orientation of the applied magnetic field with 
respect to the normal of a two-dimensional electron gas, the chemical
potential and the specific heat reveal persistent spin splitting
in all field ranges.  The corresponding shape of the thermodynamic quantities distinguishes
whether the Rashba spin-orbit interaction (RSOI), the Zeeman term or both
dominate the splitting.  The interplay of the tilting of the magnetic field and
RSOI resulted to an amplified splitting even in weak fields. 
The effects of changing the RSOI strength and the Landau level broadening are also investigated.
\PACS{
      {71.70.Di}{Landau levels}   \and
      {71.70.Ej}{Spin-orbit coupling} \and
      {73.20.At}{Electron density of states} 
     } 
} 
\authorrunning{R. Gammag and C. Villagonzalo}
\titlerunning{Persistent spin splitting of a two-dimensional electron gas}
\maketitle
\section{Introduction}
\label{intro}
New physics has been exhibited by two-dimensional electron gas (2DEG) systems even at 
weak magnetic fields. Transport properties become highly anisotropic with the direction 
of the applied magnetic field $\vec{B}$ playing a crucial role 
\cite{Eisenstein2000,Wilde2009,Studenikin2005,Jiang2009,Xia2010,Ferreira2010}. 
The anisotropy is considered to be a product of the interplay of the spin-orbit coupling, 
the Zeeman splitting and the orientation of $\vec{B}$ with respect to the 2DEG plane 
among others.  These factors determine the density of states (DOS) of the system. 
As many observables, such as the conductance are proportional to the DOS at the Fermi 
energy \cite{Winkler2003}, tuning each of these characteristics is indispensable in the design of 
devices that require the manipulation of spins.  Such devices are being eyed in 
spintronics and quantum information systems \cite{DasSarma2001}.

The spin-orbit interaction experienced by the electrons in a 2DEG can be due to the 
structure inversion asymmetric potential (Rashba) or the bulk inversion asymmetry (Dresselhaus).  Both lead to a 
zero-field spin splitting.  The former is of particular interest due to its 
versatility as it can be controlled by the gate voltage \cite{Nitta2007,Gui2004}.  Moreover, in some 
III-V \cite{Giglberger2007}, II-VI \cite{Gui2004} and Si-based \cite{Wilamowski2002} 
semiconductors the Rashba spin-orbit interaction (RSOI) normally dominates. 
Beating patterns observed in transport properties relative to the applied magnetic field were attributed to the RSOI
\cite{Wilde2009,Studenikin2005,Gui2004,Zawadzki2002}.

Tilting the magnetic field with respect to the 2DEG's normal axis has revealed
a rich variety of physical signatures.  For example, it recovers the resonant spin Hall effect
after being suppressed by an impurity scattering \cite{Jiang2009}.  Depending on the angle, 
a field's orientation can destroy some fractional quantized Hall states or turn 
anisotropic phases into isotropic ones \cite{Xia2010}.  It also causes the ringlike structures in 
the longitudinal resistivity versus magnetic field to collapse \cite{Ferreira2010}.
With such interesting tilt-induced effects, this work focuses on the influence of the applied magnetic 
field's direction on the thermodynamic properties of a 2DEG with Rashba and Zeeman interactions.  
We now investigate how a finite in-plane component of $\vec{B}$ modifies the chemical potential 
and the specific heat. The shape of their oscillations will be shown to depend on both the magnitude and orientation
of $\vec{B}$. Moreover, robust spin splitting is observed even at the weak field region.  The behavior 
of the thermodynamic quantities will be explained through the competing dominance of the Rashba 
and Zeeman interactions in various field regimes.

\section{Eigenvalues with RSOI and tilted magnetic fields}
The 2DEG considered here is lying on an $x-y$ plane and is subject to an external magnetic
field of magnitude $B = \sqrt{B_x^2 + B_y^2 + B_z^2}$.  The tilt is the angle $\theta$ that $\vec{B}$ makes with the $z$-axis.
The Hamiltonian of a single electron in this system can be written as

\begin{equation}
H = \frac{\hbar^2 \vec{k}^2}{2m^*} +  \alpha (\vec{\sigma} \times \vec{k}) \cdot \hat{z} - \vec{\mu} \cdot \vec{B},
\label{H}
\end{equation}

\noindent where the first term is the free particle energy, the second is the Rashba spin-orbit interaction, and the third is the Zeeman energy.  
The strength of the RSOI, assumed to be constant for this present work, is indicated by the parameter $\alpha$.
Here $m^*$ is the electron's effective mass, $\vec{\sigma}$ are the Pauli matrices and $\vec{k}$ is the wave vector. 
The magnitude of the latter is determined by $\vec{k} = -i \vec{\nabla} + \frac{e}{ \hbar} \vec{A}$, where $e$ is the electronic charge, $\hbar$ is 
Planck's constant over $2 \pi$, and $\vec{A}$ is the magnetic vector potential.  The magnetic moment is $\vec{\mu} = \frac{1}{2}g \mu_B \vec{\sigma}$ 
where $g$ is the Land\'{e} factor and $\mu_B$ is the Bohr magneton. Although the $g$-factor varies with $B$ \cite{Englert1982,Wang1992}, it was shown
that the $B$-dependence of the 2DEG specific heat is not significantly altered \cite{Wang1992}.  Our focus now 
is on the effects of the RSOI and tilted fields.  For this purpose, we keep $g$ constant, that is, $g = 2$ for an electron.

In solving the Schr\"odinger equation,

\begin{equation}
\left ( \begin{array}{cc}
H_n  + \Omega_z - E & i \Upsilon a + \Omega_-  \\ 
-i \Upsilon a^{\dagger} + \Omega_+   & H_n - \Omega_z - E  \end{array} 
\right) \left( \begin{array}{c} \sum_{n = 0} a_n \phi_n\\ 
\sum_{n = 0} b_n \phi_n  \end{array} \right) = 0,
\label{Eq:SE}
\end{equation}

\noindent we express the wave function solutions in terms of a superposition of the harmonic oscillator basis functions $\phi_n$ where the 
coefficients $a_n$ and $b_n$ are the unknown spin-up and spin-down complex coefficients of the $n$th Landau level, respectively.  
Here $H_n = \frac{1}{2}(a a^{\dagger} + a^{\dagger} a)$, $\Omega_j = \mu_B B_j/\hbar \omega_c$ is the $j$-th component of the rescaled Zeeman energy, 
$\Upsilon = 2 \alpha \sqrt{\zeta / \hbar \omega_c}$ is the rescaled Rashba parameter, $\zeta = m^*/2 \hbar^2$ and $\Omega_{\pm} = \Omega_x \pm i \Omega_y$.  
In this work, energy units are given in terms of the cyclotron energy $\hbar \omega_c$ where $\omega_c = eB_z/m^*$.  Equation (\ref{Eq:SE}) 
yields a set of secular equations

\begin{equation}
\left (n + \frac{1}{2} + \Omega_z - E \right) a_n + i \Upsilon \sqrt{n + 1} \, b_{n + 1} + \Omega_- b_n = 0,
\label{Eq:secular1}
\end{equation}

\begin{equation}
 -i \Upsilon \sqrt{n} \, a_{n - 1} + \Omega_+ a_n + \left(n + \frac{1}{2} - \Omega_z - E \right) b_n = 0.
\label{Eq:secular2}
\end{equation}

\noindent Note that the secular equations derived by Rashba \cite{Rashba1960} are recovered when $\vec{B}$ is normal to the 2DEG plane, that is, when we 
let $\Omega_+ = \Omega_- = 0$.

Taking advantage of the Landau level crossing, a solution can be obtained by invoking that opposite spin states of adjacent Landau levels $E_n$ are equally 
probable.  The eigenvalues $E_n^{\pm}$ can be shown to be

\begin{eqnarray}
E_n^{\pm} = n + \frac{1}{2} + \frac{\Upsilon}{2} \left(\sqrt{n+1} + \sqrt{n} \right) \nonumber\\
\pm \frac{1}{2} \sqrt{\left[\Upsilon(\sqrt{n + 1} - \sqrt{n}) + 2 \Omega_z \right]^2 + 4 \Omega_+ \Omega_-}.
\label{Eq:Evalues}
\end{eqnarray}

\noindent This result will be used to determine the density of states (DOS) which, in turn, will be used in calculating for other thermodynamic
quantities.

An important caveat has to be issued regarding (\ref{Eq:Evalues}).  It should not be evaluated for the perpendicular case $\theta = 0$.
For such a case, $\Omega_\pm$ has to be 
set to zero right at the secular equations (\ref{Eq:secular1}) and (\ref{Eq:secular2}).

The eigenvalues of equation (\ref{Eq:Evalues}) do not exhibit crossings with respect to increasing RSOI strength.  This is unlike in the 
case when $\vec{B}$ is perpendicular to the 2DEG plane. The absence of intersections was conclusively attributed to the in-plane component 
of $\vec{B}$ \cite{Bychkov1990}.  How these findings subsequently affect the chemical potential and the specific heat of a 2DEG will be 
analyzed in this paper.  It will be shown here that spin splitting in the weak $B$ region can be greatly smeared in the DOS but it can be 
clearly shown by the chemical potential and the specific heat.

\section{Chemical Potential}
\label{section:MU}
\begin{figure}
\vspace{0.64cm}
\includegraphics[width = 3.1 in]{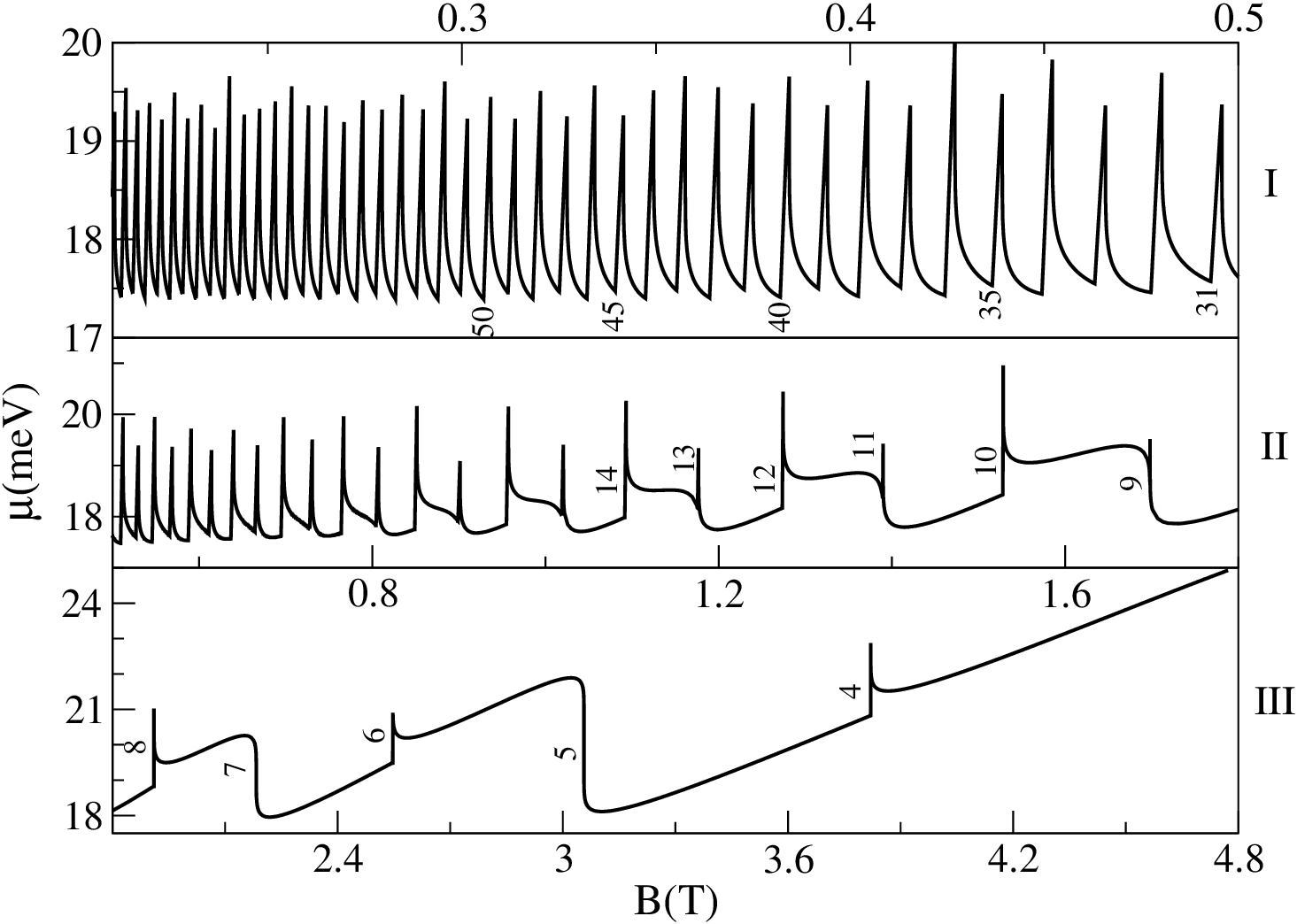}
\caption{The chemical potential as a function of the magnetic field. Here $\alpha = 1.5 \times 10^{-11}$ eV$\cdot$m, 
$\Gamma = 0.5$ meV and $\theta = 30^{\circ}$. We used $T = 10$ mK which is of the same order as in experiments \cite{Wilde2009}.
The integers coinciding with the dips indicate filling factors. The Roman numerals indicate
the different $B$ regimes as discussed in section \ref{section:MU}.}
\label{fig:muB}
\end{figure}

\begin{figure}
\vspace{0.64cm}
\centering
\includegraphics[width = 3.1 in]{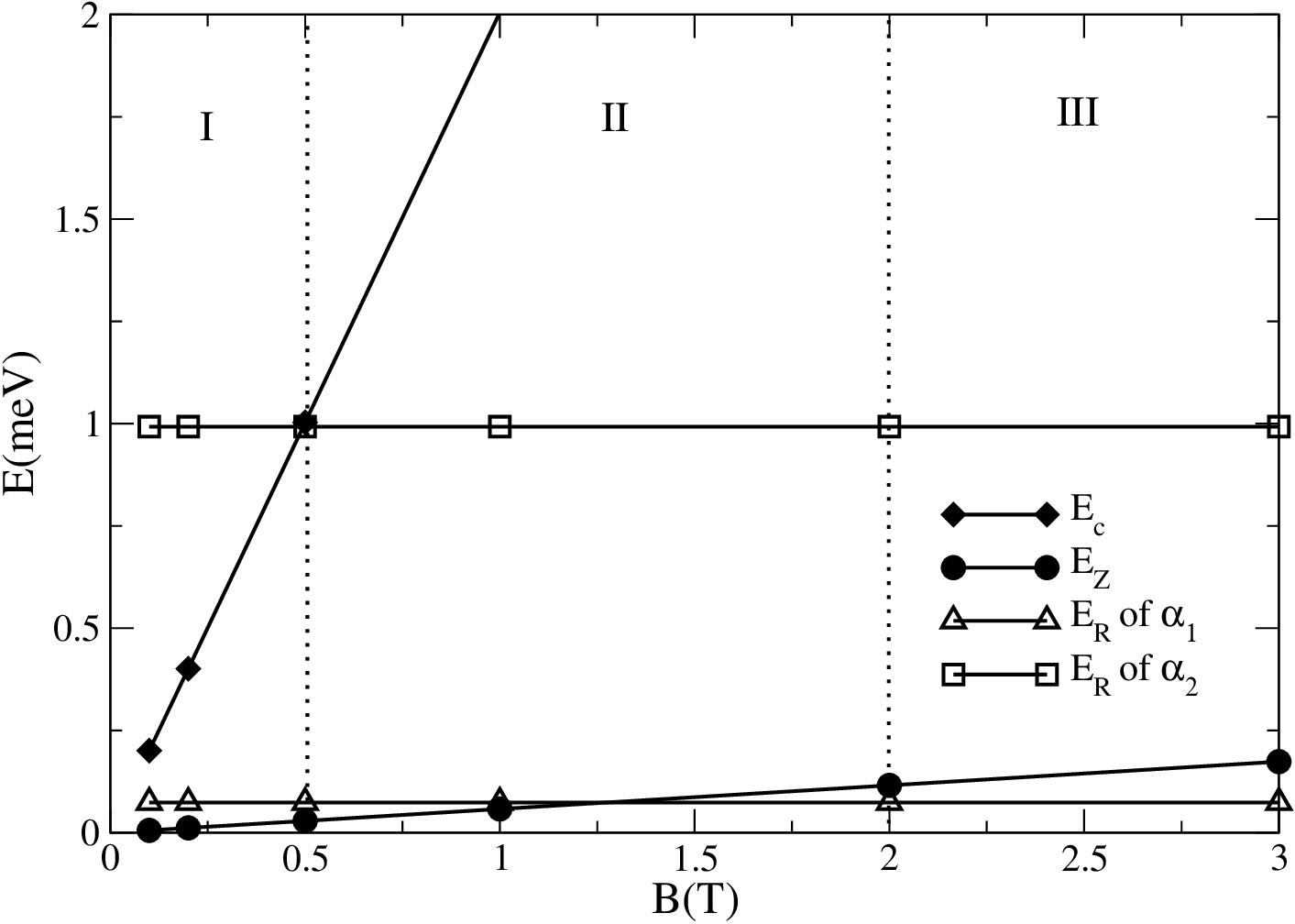}
\caption{Comparison of the different energy contributions.  Here $\theta = 30^{\circ}$, $\alpha_1 = 1.5 \times 10^{-11}$ eV$\cdot$m and
$\alpha_2 = 5.5 \times 10^{-11}$ eV$\cdot$m. The Roman numerals indicate the different $B$ regimes as discussed
in section \ref{section:MU}.}
\label{fig:EB}
\end{figure}

The chemical potential $\mu$ is the change in energy of a thermodynamic system if an additional particle is introduced, with the entropy and 
volume held fixed \cite{Chandler1987}. At zero temperature $T$, the chemical potential $\mu$ is the Fermi energy.  When $T$ is nonzero, 
$\mu$ varies with $T$ and its behavior is shown in reference \cite{VG2011} for a 2DEG in a perpendicular magnetic field.

Experiments on 2DEG devices are usually constrained to a constant particle density \cite{Studenikin2005,Wang1992,Gornik1985}.  Hence, 
to simulate conditions similar to experiments, the electron concentration is set to $N = 3.2 \times 10^{11} \mbox{cm}^{-2}$.  This is 
of the same order of $N$ in other research works \cite{Wilde2009,Studenikin2005,Xia2010,Giglberger2007,Wilamowski2002,Englert1982}.  
Knowing $N$, $\mu$ can be numerically derived from
\begin{equation}
N = \int f(E) \, \mbox{DOS}(E) \, \mbox{d}E,
\label{Eq:N}
\end{equation}
\noindent where $f(E) =  1/(\mbox{exp}[(E - \mu)/k_BT] + 1)$, $k_B$ is the Boltzmann constant and DOS($E$) is the density of states.  
Since most experiments reveal a broadenend DOS \cite{Ferreira2010,Gui2004,Englert1982,Wang1992} and theoretical fits 
\cite{Zawadzki2002,Alves2009} agree well with a Gaussian form, we also utilize it here.  The DOS can be expressed as
\begin{equation}
\mbox{DOS}(E) = \frac{eB_z}{h} \sum_n \left(\frac{1}{2\pi} \right)^{1/2} \frac{1}{\Gamma} \, \exp \left[-\frac{(E - E_n^{\pm})^2}{2 \Gamma^2} \right],
\label{Eq:DOS} 
\end{equation}
\noindent where $\Gamma$ is the broadening parameter. 
The values $\Gamma = 0.1$ meV and 0.5 meV are used in this work which are typical for numerical simulations in 2DEG \cite{Alves2009}.
In evaluating $E_n^{\pm}$ in (\ref{Eq:Evalues}), we used $m^* = 0.05 m_e$ where $m_e$ is the free electron mass.
This is the effective mass of a 2DEG in InGaAs heterostructures \cite{Nitta2007}.
For simplicity we let $B_y = 0$, $B_x= B \sin \theta$ and  $B_z = B \cos \theta$.  
Equation (\ref{Eq:N}) is solved with a maximum percent error of 10$^{-4}\%$ in $N$. 
\begin{figure}
\vspace{0.64cm}
\centering
\includegraphics[width = 3.1 in]{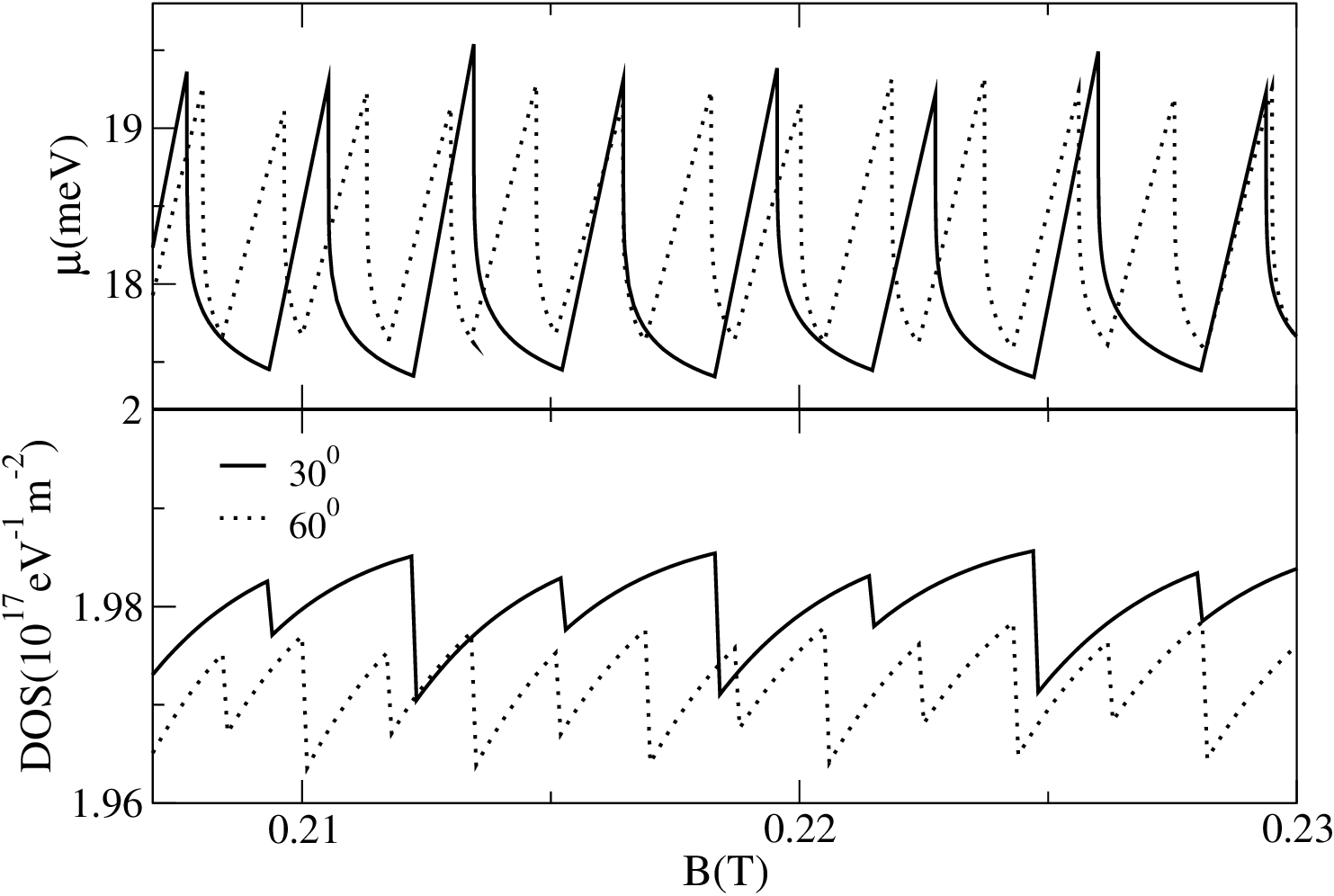}
\caption{The chemical potential for different tilt angles (top frame) and their corresponding DOS for $E = 15.8$ meV (bottom frame). Here $T = 10$ mK, 
$\Gamma = 0.5$ meV and $\alpha = 1.5 \times10^{-11}$ eV$\cdot$m.}
\label{fig:DOSTHETA}
\end{figure}
\begin{figure}
\vspace{0.64cm}
\centering
\includegraphics[width = 3.1 in]{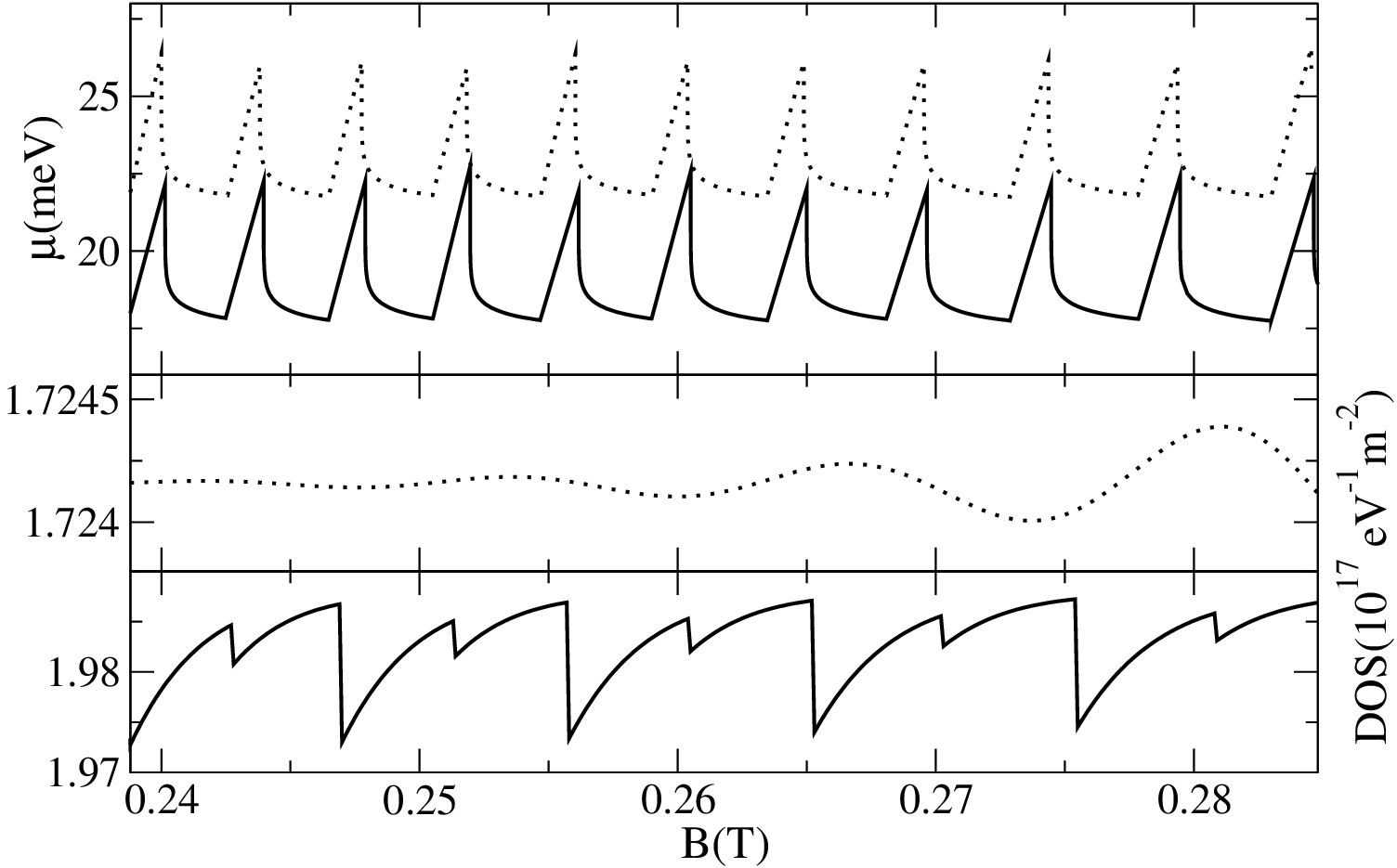}
\caption{The chemical potential for different Rashba interaction strengths (top frame) and their corresponding DOS 
for $E = 15.8$ meV (middle and bottom frame). 
Here $T =$ 4.2 K, $\Gamma = 0.5$ meV and $\theta = 30^{\circ}$.  The dotted line corresponds 
to $\alpha_2 = 5.5 \times 10^{-11}$ eV$\cdot$m 
while the solid line to $\alpha_1 = 1.5 \times 10^{-11}$ eV$\cdot$m. }
\label{fig:DOSA}
\end{figure}

In figure \ref{fig:muB}, we study the chemical potential oscillations as a function of $B$.  
The oscillations as $B$ is increased are due to the depopulation of the energy 
levels.
A good measure of the occupation of states is the filling factor $\nu$.  
It is the nominal number of filled energy levels and is inversely proportional to the component of the magnetic field normal to the 
2DEG plane, that is, $\nu = hN/eB_z$.
In the quantum Hall effect, the conductivity is quantized when $\nu$ takes on integer values.  
Because the prefactor of the DOS in (\ref{Eq:DOS}) is proportional to $B_z$,  when $B$ is increased, for a fixed $\theta$, the degeneracy of the energy 
levels also increases.  Electrons from the last occupied level will then move to lower levels until that former level becomes completely 
depopulated.  This depopulation is marked by a drop in $\mu$.
This is clearly illustrated in figure \ref{fig:muB} by the locations of the dips that fall right at integer values of the filling factor $\nu$.

It can be observed that $\mu$ behaves in three distinct ways according to its shape versus $B$ as shown in figure \ref{fig:muB}.
The description will be presented here and the next paragraphs will be devoted to its discussion.  At weak $B$ ($\nu > 50$), the oscillations 
resemble spikes of approximately uniform widths with the even and odd integer fillings having similar heights (top frame).  As $B$ increases, 
the shape of $\mu$ gradually changes.   When $B$ is moderately strong ($8 < \nu < 50$), alternating spikes appear where one is broader and 
taller and centered at even integer $\nu$ while the other is thinner and shorter and centered at odd integer $\nu$ (middle frame).  As $B$ 
further increases, $\mu$ again slowly changes shape until finally, at very strong $B$ ($\nu < 8$), large triangular waves each with a steep 
drop at odd $\nu$ develop.  The triangular waves each has a short but sharp spike that is centered at even $\nu$ (bottom frame).  The three 
regions distinguished here are, respectively, identified as follows: (I) the regime where Rashba SOI dominates, (II) the region where the 
Rashba and Zeeman interactions are comparable, and (III) where the Zeeman term prevails.  Ascribing region I (III) as RSOI (Zeeman) dominated 
is based on the fact that the RSOI (Zeeman energy) is independent (linearly dependent) on $B$.  These divisions based on the strength of 
the magnetic field are observed in all $\mu$ data obtained in this work.

When only the cyclotron energy $E_c$ is considered, that is, spin is neglected, $\mu$ peaks are large only for even $\nu$ 
\cite{GV2008,Karlhede1993,MacDonald1986}. 
The presence of odd $\nu$ peaks
for all $B$ region in figure \ref{fig:muB}, for example, is evidence of a substantial spin splitting. 
Although the Zeeman energy $E_Z$ also contributes to the appearance of odd $\nu$ peaks, it is negligible in the weak $B$ region (Region I). 
To track the contributions of the two sources of splitting,
we compare $E_c$, $E_Z$ and the RSOI energy contribution $E_R$.  Refer to figure \ref{fig:EB}.
We considered two values of the RSOI parameter $\alpha$ that are seen in literature.
We first tackle the case when the RSOI strength is equal to $\alpha_1$.    
In the region indicated by I, because $ E_c > E_R > E_Z$, the splitting is mainly contributed by RSOI.  
In region II the splitting is a combined effect of $E_R$ and $E_Z$ since the two are comparable in this $B$-range.  
Here $\mu$ behaves in a fashion where regions I and III 
shapes blend.  In Region III $E_c > E_Z > E_R$ .  Thus, the big peak of $\mu$ 
at odd $\nu$ in region III is dominated by $E_Z$.   
The latter observation is similar to the integral quantum Hall effect where the plateau of the Hall conductivity $\sigma_{xy}$ 
at even $\nu$ 
is determined by the large $E_c$ while the corresponding plateau for the odd $\nu$ is more influenced by $E_Z$ \cite{Karlhede1993}.
When stronger RSOI is considered, $\alpha_2$ for example, the spin splitting is completely dominated by RSOI and $E_Z$
can be neglected.


The distinguishing mark of figure \ref{fig:muB} in contrast with previous results \cite{Alves2009} is the appearance of a 
well-defined spin splitting in the weak $B$ case (Region I). 
This can be attributed to the RSOI plus the tilting of $\vec{B}$.  The RSOI with perpendicular $\vec{B}$ 
is normally associated with the oscillations 
in the magnetization \cite{Wilde2009,Zawadzki2002} and longitudinal resistivity \cite{Studenikin2005,Gui2004}.
They have beats 
whose amplitudes are small at weak $B$ relative to their values at high $B$. Figure \ref{fig:muB} suggests that tilting 
$\vec{B}$ amplifies the oscillations as $B \rightarrow 0$.

\begin{figure}
\vspace{0.64cm}
\centering
\includegraphics[width = 3.1 in]{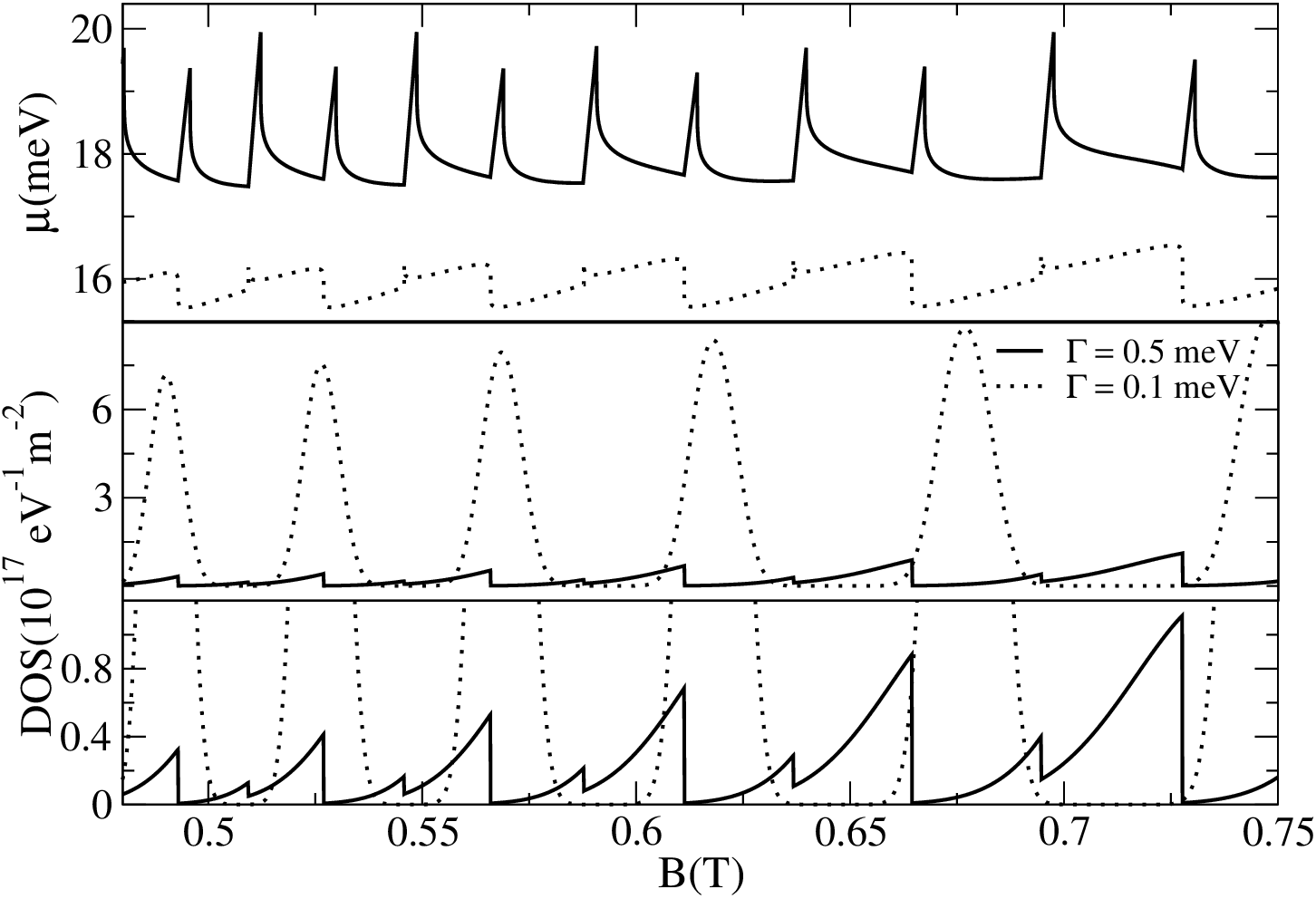}
\caption{The chemical potential for different broadening parameters (top frame) and their corresponding DOS for $E \approx 15.8$ meV 
(middle and bottom frames). To observe clearly the solid line, we magnify it in the bottom frame. Here $T = 10$ mK, 
$\theta = 30^{\circ}$ and $\alpha = 1.5 \times 10^{-11}$ eV$\cdot$m.}
\label{fig:DOSW}
\end{figure}

The sharpness of the oscillations are unexpected since we made use of a Gaussian DOS.  Refer to the lower frames of figures \ref{fig:DOSTHETA}, 
\ref{fig:DOSA}, and \ref{fig:DOSW}
which are evaluated at $E \approx 15.8$ meV. This value is chosen since it lies below but close to the Fermi energy
where the occupancy of $f(E) \approx 1$.  For $E$ below this value, we obtain similar qualitative behavior.
Closely inspecting DOS, we see the close correlation of $\mu$ and DOS. We find that although the Gaussian 
function remains the same, the interplay of the RSOI, weak $B$ and tilting effectively changed the bell shape. In this case, the contributions 
from different $E_n^{\pm}$ in the summation in (\ref{Eq:DOS}) are not uniform and at certain values of $B$, all contributions drop.

In figure \ref{fig:DOSTHETA}, the effect of varying the direction of $\vec{B}$ is displayed.  The data shown are limited to 
the weak $B$ region since all $\mu$ data 
obtained are of the same qualitative behavior as figure \ref{fig:muB}. The angle $\theta$ changes the DOS in two ways.  First is through 
$E_n^{\pm}$ which determines the location of the DOS peaks. See equation (\ref{Eq:DOS}).  
The energy levels $E_n^{\pm}$ in (\ref{Eq:Evalues}), in turn, are affected by $\theta$ through the components of the rescaled Zeeman 
energy $\Omega_j$.
We can conclude here that $\theta$ changes the frequency and phase of the $\mu$ and
DOS via the latter's dependence on the energy levels. 
Second, the DOS has a prefactor that is proportional to $B_z = B \cos \theta$.  The direction of $\vec{B}$ then affects the amplitude of 
these thermodynamic quantities because of the DOS prefactor. 

Figure \ref{fig:DOSA} shows that increasing the RSOI strength raises $\mu$.  
Here $T = 4.2$ K is chosen so that we can resolve better the 
difference between the behavior of $\mu$ for the two different RSOI strengths.
From its definition, a larger $\mu$ signifies 
that a system requires higher energy for an addition of a particle to be possible.  We can see from the figure that more 
asymmetric quantum wells demand larger amount of energy when a change in electron number is desired.  Less $\alpha$ shifts 
both DOS and $\mu$ to lower values.  However, unlike the DOS, persistent spin splitting is manifested by $\mu(B)$ data as 
$B \rightarrow 0$ (cf. dotted line in figure \ref{fig:DOSA}). The DOS spin splitting is more robust when $\alpha$ is larger but the corresponding amplitudes of the 
oscillations are shorter.  Here we can see a trade-off of broadening the Landau levels in exchange to a resolved spin splitting.    

The $\mu$ peaks in all our simulations are sharp unlike in previous results where the spin effects were ignored \cite{GV2008}.  
The oscillating behavior of the 2DEG with respect to $B$ is better resolved in $\mu$ than in the DOS as $B \rightarrow 0$.

Furthermore, we show the effect of increasing the Landau level broadening in figure \ref{fig:DOSW}.  
An increase in the broadening parameter changes the distribution of occupied states over the energy.  
Formerly unoccupied states become occupied.
This is at the expense of the once highly occupied states. Hence, the total number of states remains conserved.
A large $\Gamma$ indicates 
a system with more disorder \cite{VG2011}.  The latter can take the form of structural disorder or impurity scattering 
in which charged impurities interact with an electron via the screened Coulomb interaction \cite{Xie1990}.
Ordinarily $\Gamma$ does not drastically alter the qualitative behavior of the DOS. 
For a fixed $N$, a small $\Gamma$ is associated with a narrow DOS curve having a tall amplitude.
This is in contrast to the case of a wide $\Gamma$ where the DOS curve is broadened but its amplitude is relatively shorter. 
However, based on equation
(\ref{Eq:DOS}), the DOS has a complicated dependence on both $E$ and $B$, where the 
latter appears nontrivially in $E_n$ through equation (\ref{Eq:Evalues}).  Hence, the DOS shape in figure \ref{fig:DOSW}
is a result of the combined effects of $\Gamma$, the RSOI and the tilted field.  It is therefore difficult to attribute
conclusively to a single parameter a specific behavior of the DOS.

Despite the not-so-simple dependence of the DOS on the different parameters, 
figure \ref{fig:DOSW} suggests that a more disordered system is more resistant 
to changes in the number of particles in the 2DEG resulting to a higher $\mu$.  Here we find that even at a weak $B$, a small $\Gamma$ 
is sufficient to display the $\mu$ behavior expected only at strong fields.  This is depicted by the shape of $\mu$ for 
$\Gamma = $ 0.1 meV in the top frame of the figure \ref{fig:DOSW} which is similar to region III of figure \ref{fig:muB}. 
Comparing figures 
\ref{fig:DOSA} and \ref{fig:DOSW}, 
we find that
increasing the RSOI strength and the Landau level broadening have a similar effect on $\mu$. 
Increasing the RSOI strength increases spin splitting which causes adjacent Landau levels to overlap.
Similarly, two adjacent Landau levels will overlap if $\Gamma$ is sufficiently wide.

\section{Specific Heat}
The specific heat capacity $C_V$ is defined as the amount of energy required to raise a system's temperature by a degree while the 
volume is kept fixed \cite{Chandler1987}. It is important to determine $C_V$ so that engineers will know the heat tolerance of devices 
such as those intended to be resistant to changes in $T$.

At constant volume, $C_V$ is given as 
\begin{equation}
 C_V = \frac{\partial}{\partial T} \int f(E)\, (E - \mu) \, \mbox{DOS}(E) \, \mbox{d}E.
\label{eq:Cv}
\end{equation}

\begin{figure}
\vspace{0.64cm}
\centering
\includegraphics[width = 3.1 in]{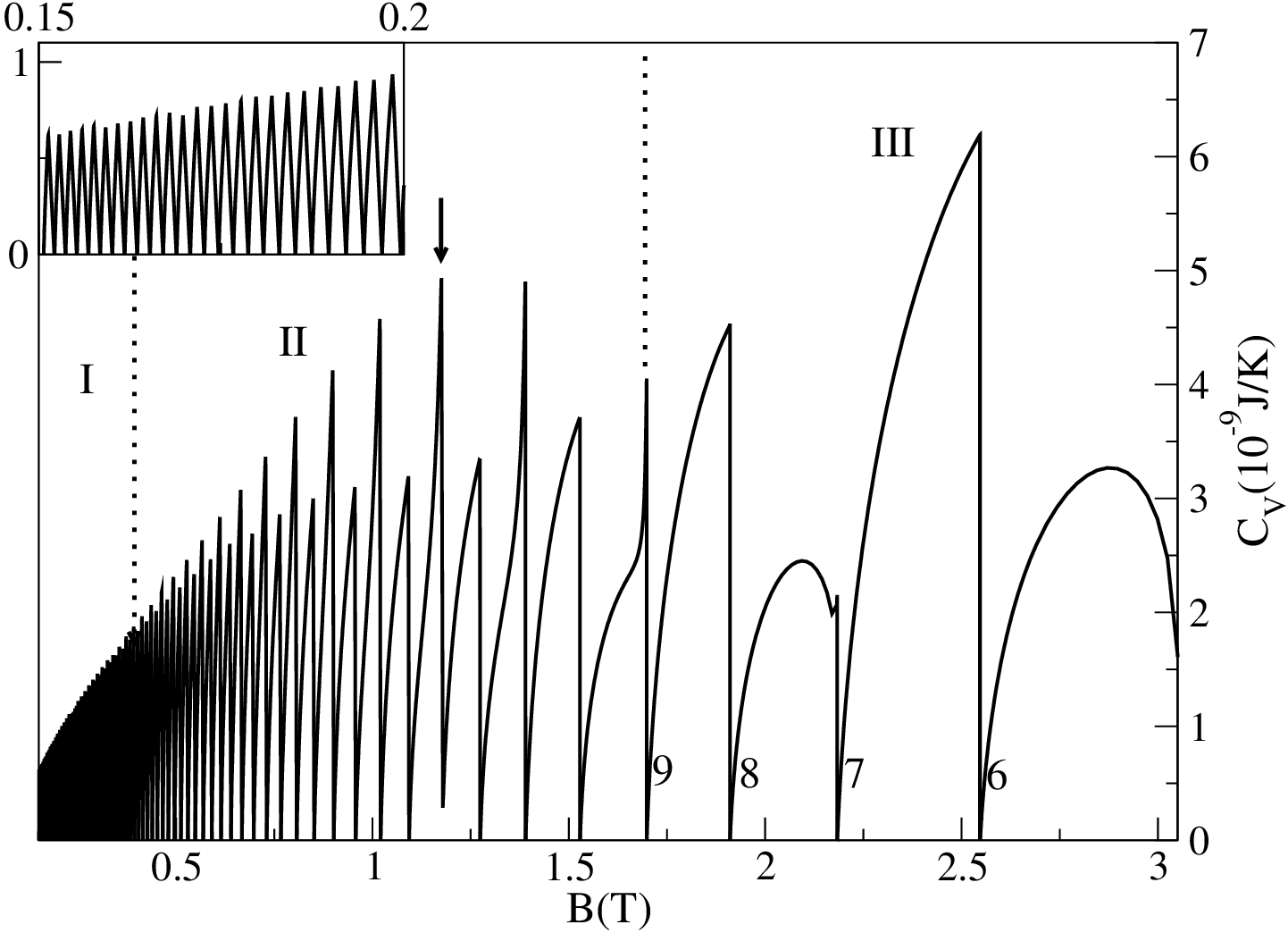}
\caption{The specific heat as a function of $B$. Here $T = 4.2$ K, $\theta = 30^{\circ}$, $\Gamma = 0.5$ meV
and $\alpha = 1.5 \times 10^{-11}$ eV$\cdot$m.  
The integers beside the dips indicate the filling factors. The inset is for the weak $B$ case.} 
\label{fig:CvB}
\end{figure}

\noindent The partial derivative in (\ref{eq:Cv}) will yield two terms:  one from $\partial f/ \partial T$ and 
another from $\partial \mu / \partial T$.  
The contribution of these two terms are taken into account in figure \ref{fig:CvB} since $\mu$
varies with both $B$ and $T$.

Experimental and numerical data show that a 2DEG under a perpendicular $\vec{B}$ has a $C_V$ that oscillates as a function of $B$ 
with the dips coinciding with integer $\nu$ \cite{Zawadzki2002,Wang1992,Gornik1985,GV2008,MacDonald1986}. These minima coinciding 
with integer $\nu$ are also demonstrated in figure \ref{fig:CvB}.  

Similar with the categories of $\mu(B)$ behavior depending on the dominant interaction, $C_V(B)$ also has three distinct features.  
In weak fields (region I), the peaks are composed of spin split levels of heights slowly increasing (see inset of figure \ref{fig:CvB}).  
When $B$ is moderately strong (region II) the oscillations are composed of peaks of alternating heights.  
Finally, when $B$ is strong, wide and tall peaks emerge with a sharp dip that cuts through each peak. 
The shape of the $C_V(B)$ at region III resembles that of the spinless case \cite{GV2008}. 
The Zeeman term in this region, creates an additional dip in the previously spin-degenerate Landau levels. 
The latter finding was also observed in earlier works \cite{MacDonald1986} where the additional dip was attributed to Coulomb interaction in
the presence of a disorder potential.

The prominence of the odd $\nu$ at weak and moderately strong $B$ gives evidence to the influence of the RSOI at these $B$ regions.
The same argument used in the preceding section applies here.  Usually only the even $\nu$ are pronounced with small dips at odd $\nu$.  
This common trend shows that the cyclotron energy $E_c$ is usually larger than any spin effect \cite{Karlhede1993}.  The comparable 
size of the peaks at strong $B$ implies that at this region, the Zeeman splitting is also comparable to the cylotron energy 
(cf. figure \ref{fig:EB}).  Both $\mu$ in figure \ref{fig:muB} and $C_V$ in figure \ref{eq:Cv} manifest an amplified spin effect in all 
$B$ spectrum.  The plots also highlight the area where one of the two spin interactions dominate or where both become comparable.

The temperature used in the simulation of figure \ref{fig:CvB} is $T = 4.2$ K, the same $T$  used in the $C_V$ measurements of Ref. \cite{Gornik1985}.  
We chose this temperature because the millikelvin range
has been found to freeze disorder effects and it constrains the system to behave as an ideal free electron gas \cite{GV2008}.  
This was also observed in reference \cite{Ramos2009} although their consideration was without RSOI.  The behavior of $C_V$ with $T$ is 
the subject of another paper.

Another notable observation is the nonmonotonic increase in the height of the $C_V$ peaks as $B$ increases where a distinct turning point 
can be found around $B \approx 1.2$ T.  See arrow pointing to $\nu = 13$ in figure \ref{fig:CvB}.  
Moreover, the dip at this point is displaced 
upward compared to $C_V \approx 0$ for the rest of the dips.  This could be reminiscent of beat nodes 
as observed in reference \cite{Zawadzki2002} 
that appears to be damped here.  
Although not shown here, we observe that the dip shifts to higher $B$ 
with $\theta$. 
This is similar to the observation that
tilting the applied field shifts the beats of the $B$-dependent magnetization \cite{Wilde2009} and 
longitudinal resistivity \cite{Studenikin2005}.

\section{Conclusions}
In this work, we have shown the distinct behavior of the chemical
potential $\mu$ and the
specific heat capacity $C_V$ of a two-dimensional electron gas subject to
a tilted magnetic
field $B$ with varying degrees of energy-level splitting interactions.
These thermodynamic properties were obtained assuming a Gaussian
density of states DOS
whose energy levels are obtained analytically in the presence of the Rashba
spin-orbit interaction RSOI
and the Zeeman effect.
The RSOI dominates at the weak $B$ region while the Zeeman energy rules
at the strong $B$ part.
Interesting $\mu$ and $C_V$ behavior occur when these two contributions are
comparable to
each other.
The combined effect of the RSOI, the Zeeman splitting and the tilting of $B$
yielded here a damping of beating patterns in the oscillations of
$C_V$ and the DOS as observed in other studies found
in literature.

Among the tunable parameters investigated, only the tilt angle $\theta$
effectively changed the phase and frequency of the $\mu$ oscillations. This outcome can be
traced to the dependence of the DOS on $\theta$ .
Finally, the interplay of the RSOI and the tilted magnetic field resulted to an
amplified spin splitting as $B \rightarrow 0$ which was not observed before.
The persistent splitting is demonstrated by the pronounced $\mu$ and $C_V$ peaks at odd filling factors
which is not observed in the spin degenerate case.
This signifies that in addition to tuning the gate voltage, in varying the RSOI
strength, the orientation of the field relative to the electron gas plane provides
control to the magnitude of the spin splitting which is important in the
spin filtering design \cite{DasSarma2001,Nitta2007} of 2DEG devices.

\section*{Acknowledgment}
R. Gammag is grateful for the Ph.D. scholarship from the Commission on Higher Education through the National Institute of Physics
as a Center of Excellence Program.

\end{document}